# A NOVEL ADAPTIVE CHANNEL ALLOCATION SCHEME TO HANDLE HANDOFFS

Alagu S[#1], Meyyappan T[*2]

[#1]Research Scholar, Department of Computer Science and Engineering
Alagappa University, Karaikudi, Tamilnadu, India
`sivaalagu@hotmail.com`
[*2]Associate Professor, Department of Computer Science and Engineering
Alagappa University, Karaikudi, TamilNadu, India
`meyslotus@yahoo.com`

*ABSTRACT*

*Wireless networking is becoming an increasingly important and popular way of providing global information access to users on the move. One of the main challenges for seamless mobility is the availability of simple and robust handoff algorithms, which allow a mobile node to roam among heterogeneous wireless networks. In this paper, the authors devise a scheme, A Novel Adaptive Channel Allocation Scheme (ACAS) where the number of guard channel(s) is adjusted automatically based on the average handoff blocking rate measured in the past certain period of time. The handoff blocking rate is controlled under the designated threshold and the new call blocking rate is minimized. The performance evaluation of the ACAS is done through simulation of nodes. The result shows that the ACAS scheme outperforms the Static Channel Allocation Scheme by controlling a hard constraint on the handoff rejection probability. The proposed scheme achieves the optimal performance by maximizing the resource utilization and adapts itself to changing traffic conditions automatically.*

## 1. INTRODUCTION

The field of wireless/mobile communication is at an interesting juncture in its development. The phenomenal worldwide growth of cellular telephony clearly demonstrates the fact that users place significant value on portability as a service feature. As end-user applications migrate towards Internet/WWW and broadband multimedia, it would be reasonable to expect strong consumer demand for wireless extensions to such services. However, this inevitable migration of mobile services to include integrated services such as data, voice, images and video represents a nontrivial architectural and technical challenge for the historically telephony-centric wireless industry. In particular, effective support to mobile users engaging in multimedia information services will require the communication path to provide sufficient buffers that cater to the change of bandwidth when the users moves from one location to another. Cellular systems deploy smaller cells in order to achieve high system capacity due to the limited spectrum. The frequency band is divided into smaller bands and those bands are reused in non-interfering cells [5].

### 1.1 Cellular Architecture

A cellular network allows cellular subscribers to wander anywhere in the region and remain connected to the Public Switched Telephone Networks (PSTN) via their wireless mobile devices. A cellular network has a hierarchical structure and it is formed by connecting Mobile Stations (MS), Base Station (BS) and Mobile Switching Centre (MSC). The Base Station serves a cell which could be few kilometers in diameter. The cell is a part of a larger region, which has been partitioned into smaller regions such that there is a base station serving each cell. All BSs within the cluster are connected to MSC. Each MSC of a cluster is then connected to the MSC of other clusters and then to PSTN. The MSC stores information about the subscribers located within the cluster and is responsible for directing calls to them [2].

Neighboring cells overlap with each other to ensure the continuity of communications when the users move from one cell to another. Certain number of channels is allocated to each base station. A channel in the system can be thought of as a fixed frequency bandwidth (FDMA), a specific time-slot within a frame (TDMA), or a particular code (CDMA), depending on the multiple access technique used. BSs and MSCs take the responsibility of allocating channel resources to mobile stations [6].

### 1.2 Handoff

Handoff (also called Handover) is the mechanism that transfers an ongoing call from one cell to another as a user moves through the coverage area of a cellular system. The handoff process is initiated by the issuing of handoff request. The power received by the MS from BS of neighboring cell exceeds the power received from the BS of the current cell by a certain amount. This is a fixed value and is called the handoff threshold. For successful handoff, a channel must be granted to handoff request before the power received by the MS reaches the receiver's threshold. The handoff area is the area where the ratio of received power levels from the current and the target BS's is between the handoff and the receiver threshold [3][7][10]. Each handoff requires network resources to reroute the call to the new base station. Minimizing the expected number of handoff minimizes the switching load. Another concern is delay. If the handoff does not occur quickly, the Quality of Service (QoS) may degrade below an acceptable level. Minimizing delay also minimizes co-channel interference. During handoff there is brief service interruption. As the frequency of these interruptions increases, the perceived QoS is reduced. The chance of dropping a call due to factors such as the availability of channels increases with the number of handoff attempts. As the rate of handoff increases, handoff algorithms need to be enhanced so that the perceived QoS does not degrade and the cost to cellular infrastructure does not increase.

## 2. CHANNEL ALLOCATION

In wireless mobile networks, the service area is divided into cells each of which is equipped with a number of channels. New originating calls in the cell coverage area and the handoff calls are sharing these channels. When any of these calls arrives at a cell where channel is not available, it has to be blocked or queued or rejected depending on the call admission control schemes. The probability of the new originating call in the cell that is rejected is called Call Rejection probability and the probability that a handoff call rejected is called Handoff Rejection Probability.

### 2.1 Handoff Decision

Generally the Handoff request is initiated either by the Mobile Station or by the Base Station. Different types of handoff decision protocols are used in various cellular systems.

#### 2.1.1 Network Controlled Handoff

In this scheme, the Mobile Switching Center (MSC) is responsible for the overall handoff decision [1]. MSC measures the signal strength and receivers threshold from different Base Stations. It then decides on the handoff request to a Base Station whose signal level is closest [1][13].

### 2.1.2 Mobile Assisted Handoff

Here the Mobile Station (MS) is responsible for finding the Base Station (BS) whose signal strength is closest to it. The MS measures the signal strengths periodically in the neighboring BS. Based on the received measurements, the BSs and MSC decides when to handoff [8][13].

### 2.1.3 Mobile controlled Handoff

In this scheme the MS has got the full control in handoff decision. Both BS and MS measures the signal strength in the neighboring BSs and the current BS sends them to the MS. The MS decides when to handoff based on the information gained from the BS and itself [8][9][13].

## 3. LITERATURE SURVEY

Many papers in the literature of related work addresses the categorization of the schemes Based on Guard channel concept. In the cellular network, channel assignment strategies can be classified into fixed, flexible and dynamic [6].The existing literature addresses the Static Guard Channel allocation exclusively for handoff and fixed channel system where there are no separate guard channels exclusively for handoff. In fixed channel assignment (FCA) scheme, fixed numbers of channels are assigned to each cell and there isn't any Guard Channel set aside for handoff requests only. Whenever new call request or handoff request arrives, the base station will check to see if there is a channel available in current cell. The call will be connected if there is a channel available and it will be dropped if there isn't any channel left. So handoff request and new call request are dealt with equally. The cell doesn't consider the difference between Handoff request and new call request. It assigns the channels to BS by First Come First Serve basis [4][11][12]. The Quality of Service is not satisfied because the handoff blocking rate is as same as new call blocking rate. The so called "Guard-channel" (GC) concept offers a means of improving the probability of a successful handoff by reserving a certain number of channels allocated exclusively for handoff requests. The remaining channels can be shared equally between handoff requests and new calls [1][8]. Allocating Guard channels for Handoff improves the overall throughput which was discussed in our previous papers [11][12]. If the guard channel number is too big, the new call blocking rate will be high because several channels are set aside for handoff requests even when the traffic load is low. In this case, the resources are wasted by not serving either for handoff request or new call request. If the number is too small, the handoff blocking rate can't be guaranteed under high traffic load. So this scheme enhances the QoS by reducing the handoff blocking rate in a stable traffic load. While when the traffic load is changing periodically or dynamically due to big event or working rush hours, it is not flexible enough to get good QoS.

## 4. PROPOSED WORK

In this paper, the authors devise a scheme Adaptive Channel Allocation Scheme (ACAS) in which the channels for handoff requests are dynamically allocated based on the observation of certain past period in the network. This scheme is aimed to utilize the available resources efficiently and also to balance the load in the network traffic.

### 4.1 Efficient Resource Utilization

The new call dropping rate determines the fraction of new calls that are rejected, while the handoff blocking rate is closely related to the fraction of admitted calls that terminate prematurely due to handoff. The resource utilization is the efficiency of use of the limited channels. Maximum resource utilization is the main objective of this research work. For example, to get good resource utilization, less number of Guard channels is assigned to the handoff calls when the number of handoff requests is less under the low traffic load. If more channels are saved for the handoff request in this condition, the resources are wasted because the channels don't serve either for handoff request or new call request. At the same time if the number of handoff requests is high then more number of Guard channels is required to handle it. The balance of the new call rejection rate and handoff call rejection rate are monitored and maintained to get better resource utilization in cellular network.

### 4.2 Adaptive Channel Allocation Scheme (ACAS)

The selection of number of guard channels exclusively for handoff call is essentially important factor to get good Quality of Service. For different type of traffic load and mobility factor, different number of guard channels is needed to be allocated. The number of guard channels can't be fixed when the traffic load is changing with the time. The authors addresses this problem through the proposed scheme ACAS.

The Adaptive Channel Allocation Scheme automatically searches the optimal number of Guard Channels to be reserved for handoff calls at each BS. In this paper the author considers for a Base Station BS, having total number of channels C, the Guard channels exclusively for handoff will be GCh. The rest of the available channels are used by the new originating calls in that cell and also by the handoff calls, which is say Co. A new call request will be granted for admission if the total number of on-going calls (including handoff calls from other cells) is less than the number Co. A handoff call request will be granted for admission if the total number of on-going calls in the cell is less than the total capacity C.

The algorithm ACAS can be illustrated as follows:

**Data Structure**

Consider the following parameters in a particular BS coverage area,

    The total number of available channels                – C

    Open Access Channels (new calls + Handoff calls)    – Co

    Guard channels for handoff calls                    – GCh

    Where, C = Co + GCh, Co = C – Gch and GCh is allocated dynamically

Oc = number of on-going calls

Nc = number of admitted new originating calls

Hc = number of admitted handoff calls

H = Total number of handoff call(admitted+rejected)

Where, Oc = Nc + Hc

Rn = number of rejected new originating calls

Rh = number of rejected handoff calls

t = time period

Th = Threshold for handoff call rejection probability

Algorithm: ACAS (t,C ) // the algorithm takes time period and channels as input

{

Co=C-GCh

For every handoff call request Do

{

  If Oc < C, then

   {

    Hc = Hc + 1 and grant admission

    Oc = Oc +1

   }

Otherwise, Rh = Rh +1 and reject.

}

For every new call request Do

{

  If Oc < Co, then

  {

   Nc=Nc+1 and grant admission

   Oc = Oc +1

  }

  Otherwise, Rn = Rn +1 and reject.

}

If a call is completed or handoff to another cell

{

Oc = Oc – 1

Check with MSC whether the ended call is handoff call or new originated call

If handoff call then Hc = Hc-1

Else Nc = Nc-1

}

If a handoff call is dropped and Rh/H >= AuTh then

{

GCh = min{ GCh +1, Cmax}

If Rh/H <= AdTh for N consecutive handoff calls, then

GCh = max{GCh – 1, Cmin}

}

Nc,Hc are reported to monitor the successful handoff and new calls at a specified time period.

}//end of the algorithm: ACAS

As the number of Guard Channels allotted plays a vital role to the key performance, it is dynamically altered every specific time period say t. In this approach the number of guard channels which is to be allocated is determined through optimizing certain performance goal with service quality constraints. When a base station experiences high handoff blocking rate, the number of guard channels will be increased until the handoff blocking rate drops to below its threshold. When a base station does not get to use a significant portion of the guard channels over a period of time, the number of guard channels is gradually decreased until most of the guard channels are used frequently. By doing this, the handoff blocking rate is controlled to close to its threshold.

The proposed algorithm increases the number of guard channels when a handoff call is dropped under the condition that $Rh/H >= Au*Th$, and it decreases the number of guard channels after a number of consecutive handoff calls under the condition that $Rh/H <= Ad*Th$. Au and Ad are usually chosen to be less than 1. By choosing $Au < 1$, the algorithm will most likely keep the handoff blocking rate below its given threshold.

The simulation studies are performed for comparisons of the proposed algorithm with fixed channel allocation (FCA) and static guard channel allocation policy. The result proves that the algorithm guarantees that the handoff dropping rate is below its given threshold and at the same time the new call dropping rate is minimized.

## 5. EXPERIMENTAL RESULTS

The author simulates the algorithm in a six cells as a part of full network. The simulation program is implemented in Turbo C++, version 3.0, and run under MS DOS 6.2 environment. Object oriented approach is used to implement the real world environment. Results are directed to a text file and graph for the same is obtained using MATLAB. The result comprises of comparison between the 3 schemes as Handoff Handling without using Guard Channels-FCA, Using static Guard Channels, using Adaptive Guard Channels (ACAS). The following are the initial parameters chosen for simulation.

$Oc=0, Nc=0, Hc=0, Rn=0, Rh=0, H=0$

t=10 seconds // time period for updating the measurements

C=20 // No. of channels

GCh=10

Au=0.9

An=0.6

Th=0.2

N=10 // No. of consecutive calls

The following graphs show the comparative study of the three schemes, Fixed Channel Assignment without using Guard Channels, Static Guard channel assignment and Adaptive Guard channel assignment.

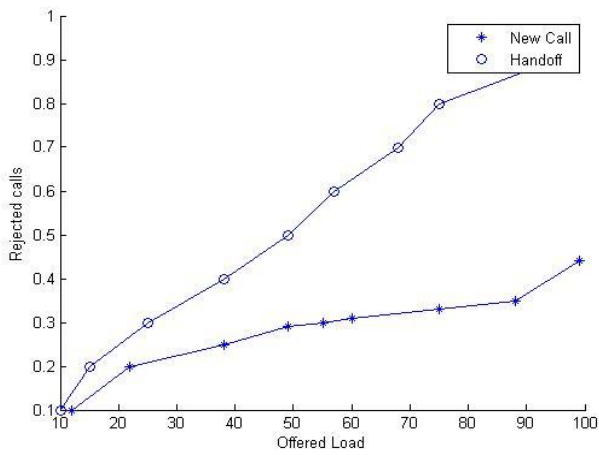

**Figure 1. Fixed Channel Assignment (FCA) strategy without using Guard Channels**

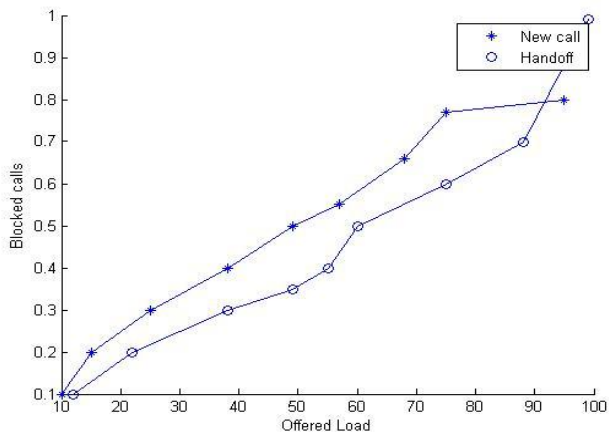

**Figure 2 Static allocation of Guard Channels exclusively for Handoffs**

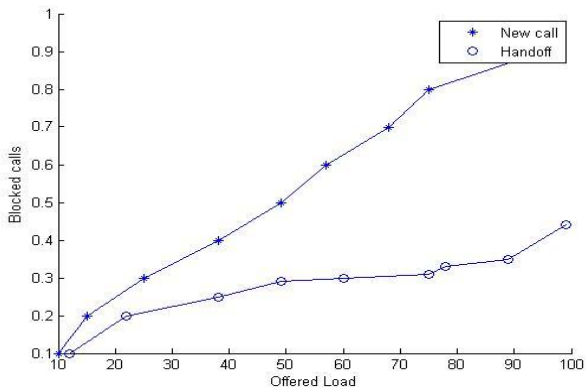

**Figure 3. Adaptive Channel Allocation Scheme**

Figure 1 shows the simulated output of the FCA scheme where no guard channels are allocated for handoff calls. It simply works in FCFS (First Come First Served) manner. The output graph shows that the Handoff Call rejection rate is slightly high than the new originating call rejection rate. But it should be noted here that handoff calls should be given higher priority than the new originating calls. The overall performance is also not satisfactory since both new call rejection and handoff call rejection ratio is comparatively high.

Figure 2 shows the simulated output of the static guard channel allocation scheme i.e., fixed number of guard channels exclusively allocated for handoff. Here the number of handoff call rejection is reduced but the new call rejection is highly increased because the number of guard channels allocated is high than which is actually required. Moreover in some cases if the number of guard channels is less, then handoff rejection rate will increase and hence affect the throughput.

Figure 3 shows the simulated output of our proposed scheme - Adaptive Channel Allocation Scheme (ACAS). Here the channels are not allocated static and they are allocated based on the traffic in the past certain period of time. The number of guard channels gets dynamically adjusted and it is clearly seen from the graph that both new calls and handoff calls utilizes the channel properly and the call rejection rate is low for both. Hence there is tradeoff.

## 6. CONCLUSION

In this paper, the authors presented a new way of Adaptive channel allocation scheme (ACAS) in cellular networks. The aim of the scheme is to effectively utilize the available resources. The main problem faced in Guard channel allocation is the number of Guard channels chosen. The proposed algorithm adjusts the number of guard channels dynamically according to the dropping rate of handoff calls in a certain period of time. It tries to make sure that the handoff call rejection rate is below the given threshold and it also tries to reduce the new call rejection rate by decrementing the number of guard channels when it is observed to be more than needed. The proposed scheme – ACAS has high degree of spectrum utilization with good QoS.

**Authors**

**Mrs.S.Alagu, M.Sc., M.Phil.,** currently Ph.D. Research Scholar in Department of Computer Science and Engineering, Alagappa University, Karaikudi. She has a teaching experience of 10 years. She has published research papers in National and International Journals and Conferences. Her research area includes Wireless Mobile Networks.

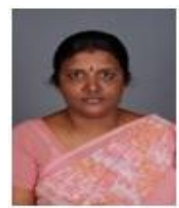

**Dr. T. Meyyappan M.Sc., M.Phil., M.B.A., Ph.D.,** currently, Associate Professor, Department of Computer Science and Engineering, Alagappa University, Karaikudi, TamilNadu. He has obtained his Ph.D. in Computer Science in January 2011 and published a number of research papers in National and International journals and conferences. He has been honored with Best Citizens of India Award 2012 by International Publishing House, New Delhi. He has developed Software packages for Examination, Admission Processing and official Website of Alagappa University. His research areas include Operational Research, Digital Image Processing, Fault Tolerant computing, Network security and Data Mining.

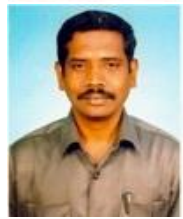